\documentclass{PoS}

\title{Deflation Methods in Fermion Inverters\thanks{This work was partially supported by the National Science Foundation, Computational Mathematics Program, under grant 0310573.}}

\ShortTitle{Deflation Methods in Fermion Inverters}

\author{\speaker{Walter Wilcox}\\
        Department of Physics, Baylor University, Waco, TX, USA 76798-7316\\
        E-mail: \email{walter\_wilcox@baylor.edu}}

\abstract{I will review recent developments in matrix deflation methods,
by Ronald Morgan/Walter Wilcox, Andreas Stathopoulos/Konstantinos Orginos, and Martin L\"uscher,
with application to lattice QCD fermion inversion. I will begin with a short review of deflation-related work in the field. The Morgan/Wilcox algorithms using GMRES and BiCGStab for deflation will be described. Typical results for quenched (Wilson and twisted mass) and dynamical configurations (CP-PACS and ETMC) will be displayed and discussed. I will outline how inclusion of multiple right-hand sides and multiple shifts can be accommodated within versions of these methods. I will also review deflation methods based on Conjugate Gradient introduced by Stathopoulos and Orginos for hermitian systems with multiple right-hand sides as well as L\"uscher's non-hermitian domain decomposition algorithm using Generalized Conjugate Residual.}

\FullConference{The XXV International Symposium on Lattice Field Theory\\
		 July 30 - August 4 2007\\
		 Regensburg, Germany}

\begin{document}

\section{Introduction and previous work}

What is matrix deflation? A good, working definition is that it consists of the effective removal or projection of the low lying eigenmodes of a matrix in order to speed up additional calculations with the same matrix. It is known, for Hermitian systems, that the convergence rate for numerically solving linear equations is proportional to the condition number, which is defined as the square root of the ratio of the largest to the smallest eigenvalue, $\sqrt{\lambda_{n}/\lambda_{1}}$. Reducing this ratio will effectively speed up the solution of the associated linear equations. For lattice QCD these small eigenvalues occur at small quark masses, and the general numerical problem being encountered is that of critical slow-down.

The first discussion I am aware of where deflation is used in a lattice QCD calculation was in the plenary talk of Ref.~\cite{lbl1}, which reported some computer experiments for system of equations for multiple right-hand sides. Unfortunately, the results did not show great promise then because the eigenvalues of the systems studied were not sufficiently small. The application and use of a separate deflation step in the acceleration of the inner loop of overlap fermions, which are Hermitian shifted systems, is documented in Refs.~\cite{lbl2} and \cite{lbl3}. A number of authors, including those of Ref.~\cite{lbl4}, have used deflation techniques in the context of calculating flavor-singlet propagators using the Hermitian Wilson-Dirac operator. See also Ref.~\cite{lbl5} for a similar method using staggered fermions and Ref.~\cite{lbl6} for a small review of such techniques. A technique for speeding up the calculation of meson two-point functions by averaging the low-lying eigenmodes over all spatial positions of the quarks was introduced in Ref.~\cite{lbl8}; a similar method was also introduced in Ref.~\cite{lbl9}. In addition, low-mode preconditioning in the so-called $\epsilon$-regime was studied in Ref.~\cite{lbl7}.

I will now describe some of the recent approaches used to implement deflation in the context of lattice QCD fermion inverters, concentrating on the methods developed in the context of Morgan and Wilcox's general non-hermitian deflation~\cite{pap1,pap2}, Stathopoulos' and Orginos' hermitian deflation~\cite{lbl12}, and L\"uscher's domain decomposition deflation~\cite{Lusch}. Refs.~\cite{PoS1} and \cite{SandOProc}, discussing the Morgan/Wilcox and Stathopoulos/Orginos algorithms, respectively, are included in these Proceedings. Preprints and talks given at this conference using deflation or related techniques are listed in Refs.~\cite{lbl10,lbl11}. 

\section{Morgan's and Wilcox's work on non-hermitian deflation}

In the system of equations, $Ax=b$, where $A$ is the known $n\times n$ matrix,  Krylov methods like GMRES and CG develop a polynomial solution in powers of $A$ times $r_0$, where $r_0$ is the residual vector. One can show the exact polynomial should have a zero at each eigenvalue in the complex plane. After $m$ iterations, the Krylov subspace, $K$, is given by
\begin{equation}
K=Span\{r_0, Ar_0,  A^2r_0,... A^{m-1}r_0\}.
\end{equation}
The polynomial developed in this space has value unity at the origin in the (possibly complex) eigenvalue space. Therefore, for low $m$ values it is necessary for this polynomial to make a sharp turn to zero near the origin if the system has small eigenvalues; this is why it is difficult for restarted methods to develop the correct small eigenvalue spectrum. Once the small eigenvalue spectrum has been developed, it is time-consuming and wasteful to re-develop the same small eigenvalue spectrum for multiple uses of the matrix, such as occurs when multiple right-hand sides are solved. It would be better to try to effectively remove or deflate such eigenvalues, in the sense of the above introduction, to produce a more efficient simulation. This is the basic numerical problem faced by users of fermion matrix inverters in lattice QCD at small quark masses.

Deflation has been studied in the context of restarted
GMRES~\cite{GMRES-E,KhYe,ErBuPo,ChSa,Sa95B,BaCaGoRe,BuEr,LCMo,DS99,GMRES-IR,
GMRES-DR}. One of these approaches is related to Sorensen's
implicitly restarted Arnoldi method for eigenvalues~\cite{So} and is called
GMRES with implicit restarting~\cite{GMRES-IR}. A mathematically
equivalent method, called GMRES with deflated restarting
(GMRES-DR)~\cite{GMRES-DR}, is also related to Wu and Simon's restarted
Arnoldi eigenvalue method~\cite{WuSi}, and is the method qualitatively described in this section. The Morgan/Wilcox work on deflation began in Ref.~\cite{NP1} in the context of GMRES, and was followed later by Ref.~\cite{NP2}, which discusses incorporating multiple shifts and multiple right hand sides for GMRES and a deflated version of BiCGStab. The detailed description of these algorithms is given in \cite{realfirst,pap1,pap2}.

An important feature of GMRES-DR is that it computes eigenvalues and solves
linear equations simultaneously. Approximate eigenvectors corresponding to the small 
eigenvalues are computed at the end of each cycle and are put at the beginning of the next
subspace. Letting $r_0$ be the initial residual for the linear equations
at the start of the new cycle and $\tilde y_1, \ldots
\tilde y_k$ be harmonic Ritz vectors~\cite{IE,Fr92,PaPavdV,IEN}, the
subspace of dimension $m$ used for the new cycle of GMRES-DR($m,k$) is
\begin{equation}
Span\{\tilde y_1, \tilde y_2, \ldots \tilde y_k, r_0, A r_0, A^2 r_0, A^3
r_0, \ldots , A^{m-k-1} r_0 \}. \label{ss}
\end{equation}
This can be viewed as a Krylov subspace generated with starting vector
$r_0$ augmented with approximate eigenvectors. Remarkably, the whole
subspace turns out to be a Krylov subspace itself (though not with $r_0$ as
starting vector)~\cite{GMRES-IR}. Once the approximate eigenvectors are moderately accurate, their inclusion in the subspace for GMRES essentially deflates the corresponding
eigenvalues from the linear equations problem. 
\begin{figure}
\begin{center}
\leavevmode
\includegraphics*[viewport=0 0 800 550, scale=0.46]{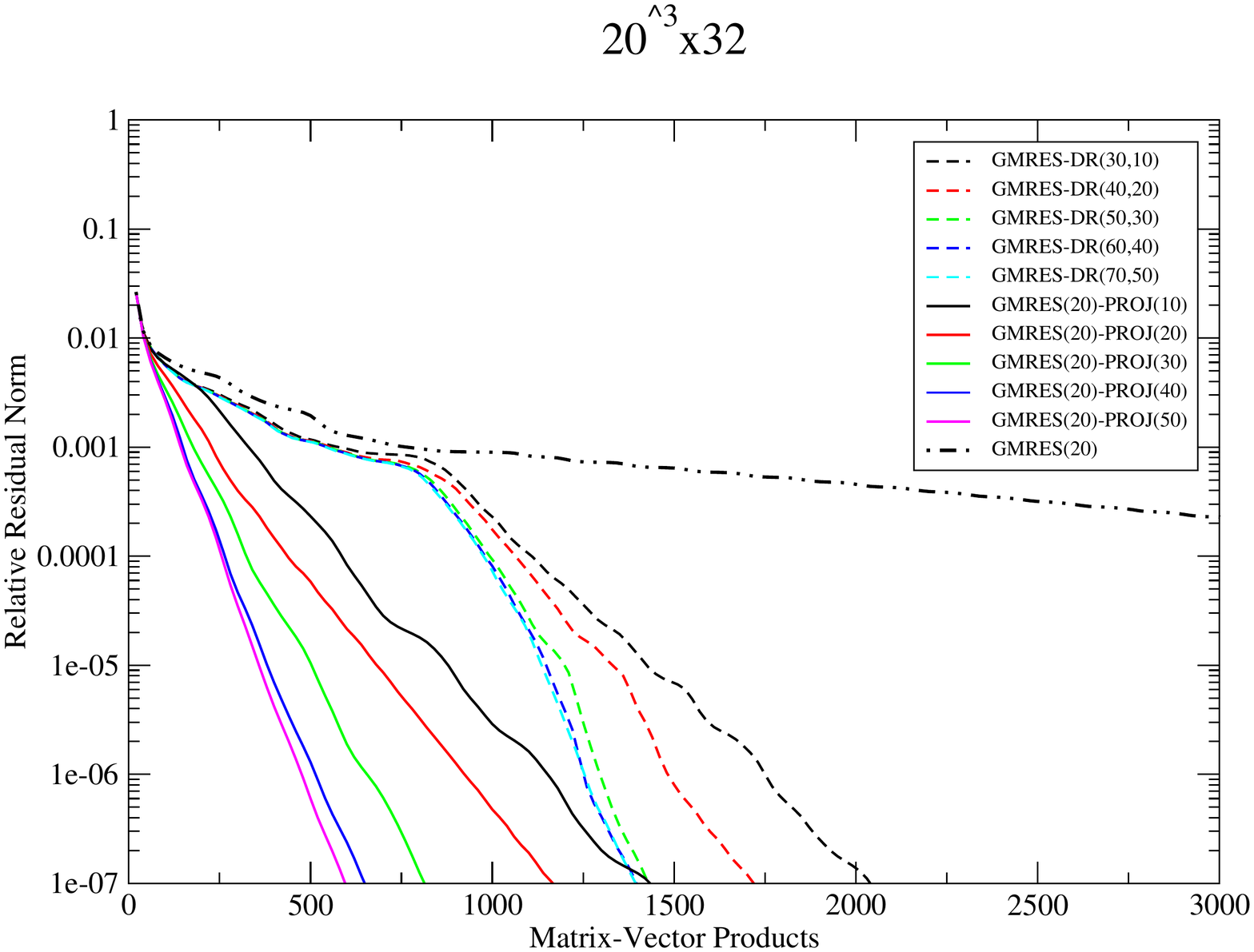}
\caption{The residual norm as a function of matrix vector products for various solvers on a quenched $20^3\times 32$ Wilson configuration at $\kappa_{cr}$.}
\vspace{1cm}
\includegraphics*[viewport=0 0 800 550, scale=0.46]{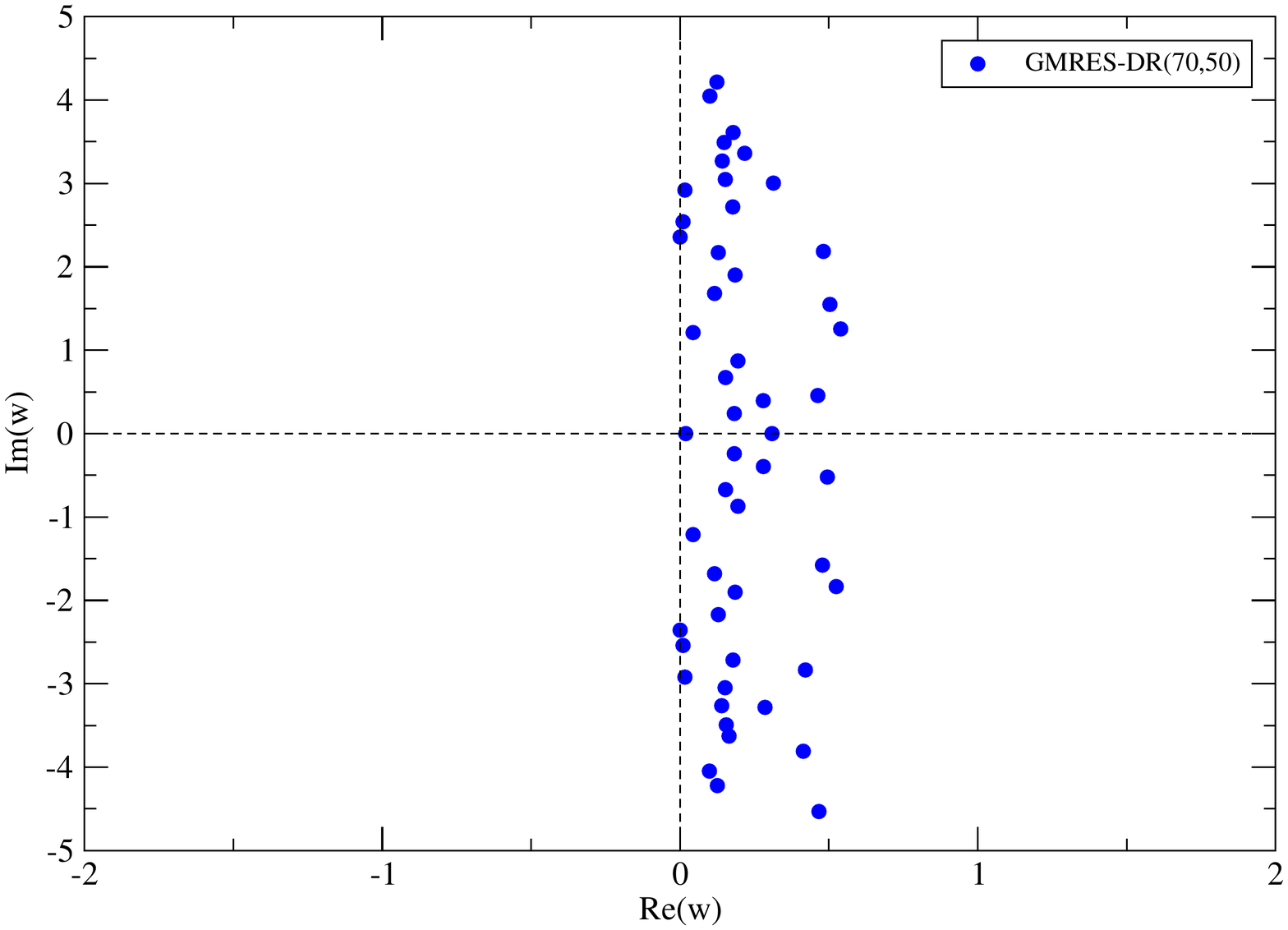}
\caption{The approximate eigenspectrum associated with the matrix used in Fig.~1 calculated by GMRES-DR(70,50) with a residual norm stopping criterion of $\approx 10^{-8}$.}
\end{center}
\end{figure}

\begin{figure}
\begin{center}
\leavevmode
\includegraphics[scale=0.46]{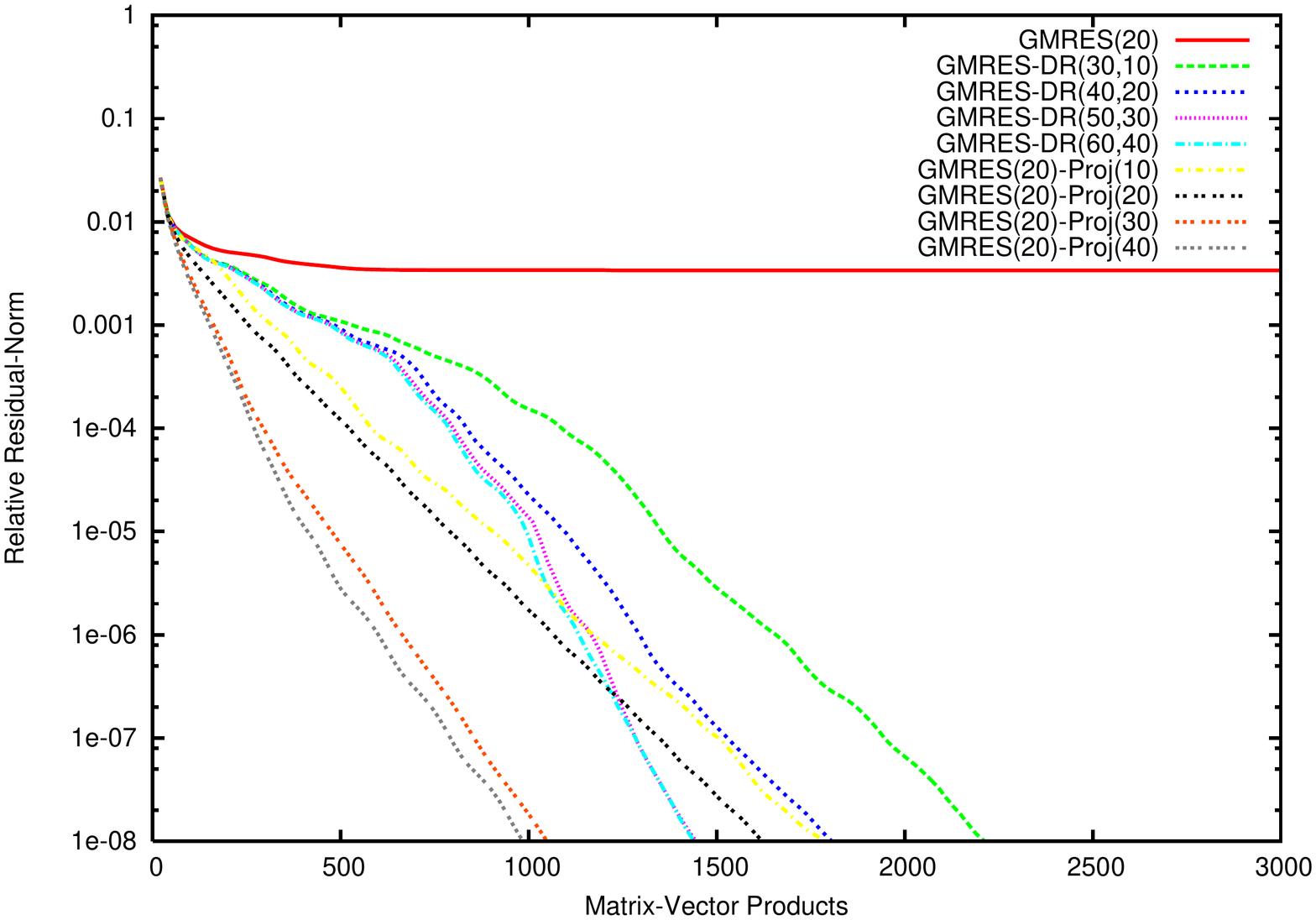}
\caption{The residual norm as a function of matrix vector products for various solvers on a quenched $20^3\times 32$ twisted mass configuration.}
\end{center}
\end{figure}

The approximate eigenvectors in GMRES-DR span a small Krylov subspace. They are generated in a compact form by
\begin{equation}
 AV_k = V_{k+1} \bar H_k, \label{recur1}\label{AV}
\end{equation} 
where $V_k$ is a $n$ by $k$ orthonormal matrix, $V_{k+1}$ is the same except for an extra column, and $\bar H_k$ is a full $k+1$ by $k$ matrix. The columns of $V_k$ span the subspace of approximate eigenvectors. At the end of a cycle of GMRES-DR, the harmonic Ritz values are computed. The matrix $V_k$ is then formed so that it is orthonormal and has columns spanning the harmonic Ritz vectors corresponding to desired harmonic Ritz values. Note this compact form is similar to an Arnoldi recurrence, and it allows access to both the approximate eigenvectors and their products with $A$ while requiring storage of only $k+1$ vectors of length $n$.  

Using the eigenvector information generated by GMRES-DR($m,k$), subsequent systems
are solved by combining restarted GMRES with a projection between cycles over the previously determined eigenvectors. The second and subsequent right-hand sides are solved much quicker than without the deflation. We call this method GMRES($m$)-Proj($k$), where \lq\lq $k$" is the number of deflated eigenvectors.

An example of the use of GMRES-DR followed by GMRES-Proj is given in Fig.~1. This system consists of a $20^3\times 32$ quenched Wilson gauge configuration, evaluated essentially at $\kappa_{cr}$. (In this and the following, we actually solve the even/odd preconditioned system.) In order to make the problem as realistic but difficult as possible, we have chosen a $\kappa$ in each problem which makes the real part of the lowest Ritz eigenvalues essentially zero. The value used for this configuration, $\kappa=0.15720$, is very close to $\kappa_{cr}=0.15691$~\cite{Jan1}. Fig.~1 shows that GMRES(20) does not converge after 3000 iterations. It is incapable of forming the low eigenvector polynomial because of the small restart value. On the other hand, the various GMRES-DR residual norms show clearly that the correct small eigenvalues in the spectrum, seen in Fig.~2, are understood and dealt with efficiently. The evidence for this is the development of the sometimes sharp "knee" in the rate of convergence of the residual vector as a function of matrix vector products (MVPs). This phenomenon is called "super-linear convergence" and is seen near 1000 MVPs in this example. The various larger ($m,k$) pairs have a knee at approximately the same number of MVPs, but make an increasingly faster convergence after this point. The eigenvalues of each GMRES-DR($m,k$) run are then passed on to GMRES($m-k$)-Proj($k$). For exact comparison, we actually use the same right hand side in this and the other examples in this paper. Notice that the residual vector slopes of the GMRES-DR($m,k$) and the corresponding GMRES($m-k$)-Proj($k$) runs, after super-linear convergence, are approximately parallel, and that this slope is applied essentially from the beginning of the GMRES-Proj run, leading to a significantly faster congergence of the system. Also notice that there is only a small change in the residual curves for both GMRES-DR and GMRES-Proj when going from 40 to 50 deflated eigenvalues. We say the residual \lq\lq saturates" at this point.

The approximate low eigenvalue spectrum of this system, calculated with the initial run of GMRES-DR(70,50) is shown in Fig.~2. One can see the approximate positive/negative imaginary value symmetry in the spectrum as well as the fact that a single real eigenvalue appears very close to the origin, making the problem tough for any method. The stopping criterion on the residual norm on this run was $\approx 10^{-8}$, which would be considered full convergence on the linear equation system. However, for increased convergence of GMRES-Proj on multiple right-hand sides it is often useful to run the GMRES-DR well past the point where the linear equations have converged.

\begin{figure}
\begin{center}
\leavevmode
\includegraphics[scale=0.62]{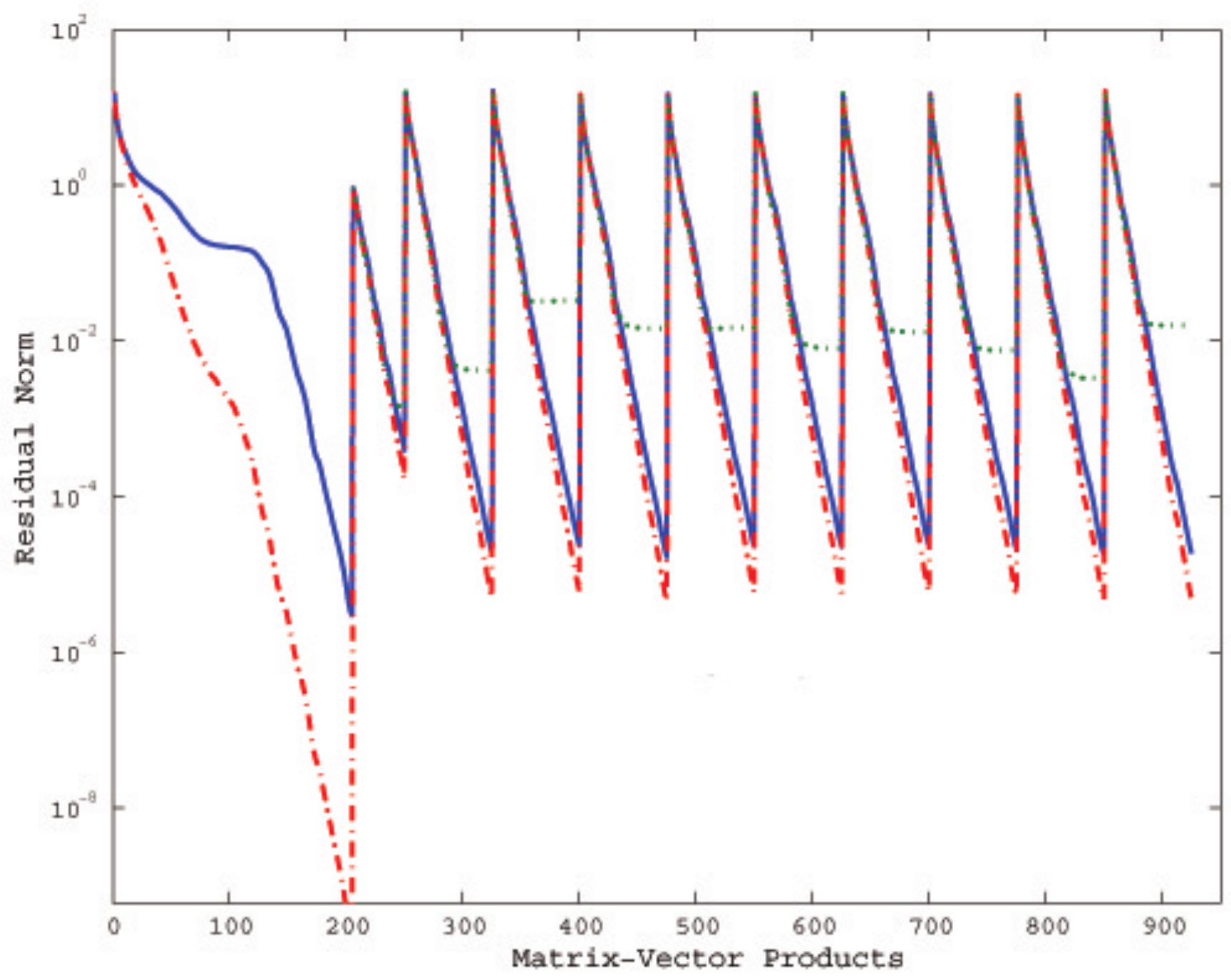}
\caption{Illustration of shifting while solving 10 additional right-hand sides for a MATLAB matrix. The non-hermitian matrix used was of order 2000.}
\vspace{1cm}
\includegraphics[scale=0.58]{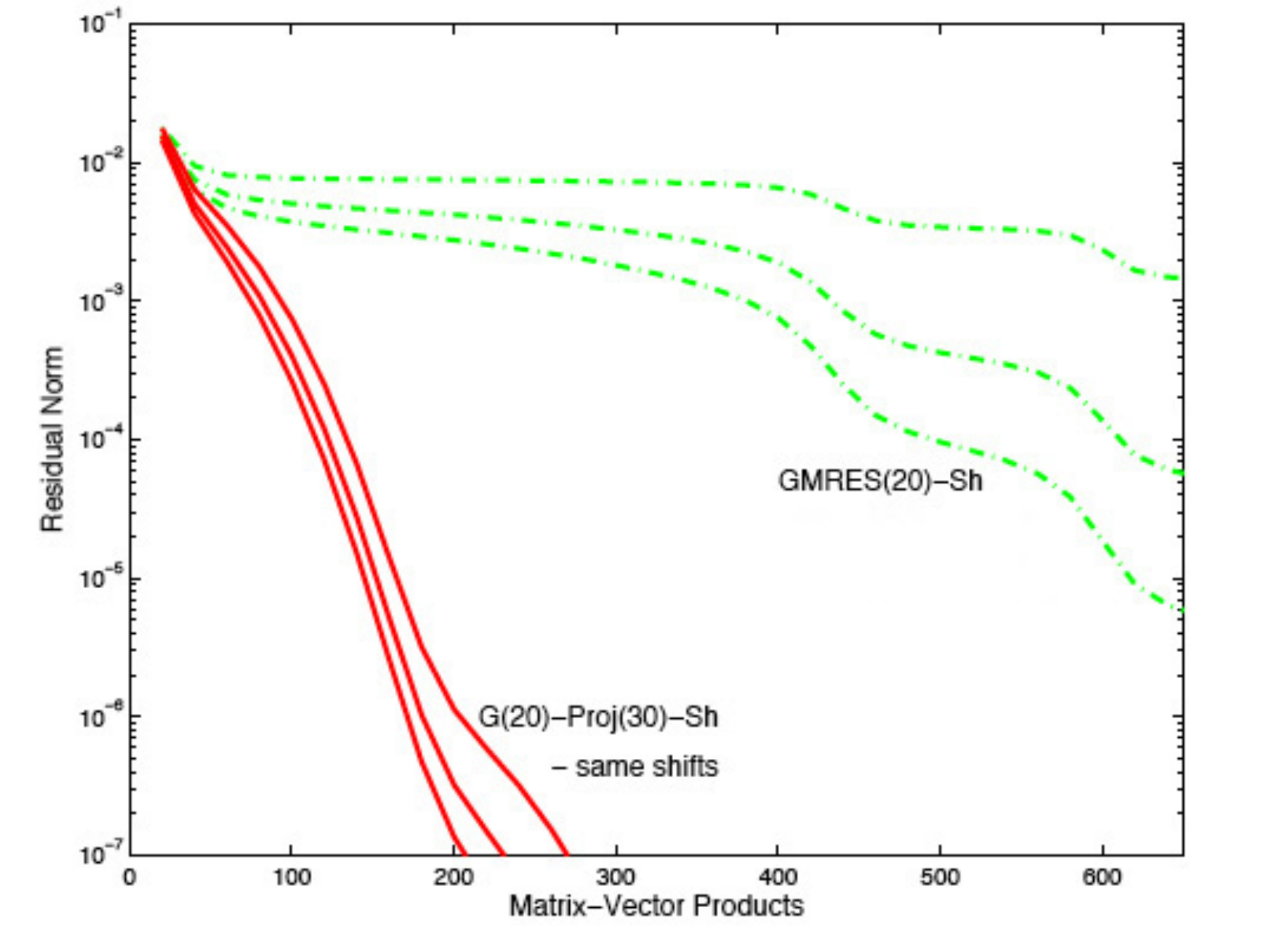}
\caption{Illustration of shifting while projecting using GMRES-Proj-Sh on a new right-hand side for a $16^4$ Wilson matrix.}
\end{center}
\end{figure}

Results for GMRES-DR and GMRES-Proj on a $20^3\times 32$ twisted mass configuration, at $\kappa=0.15728$, $\mu=0.005$ are shown in Fig.~3. (This is the lowest maximally twisted quark mass considered in Ref.~\cite{AbdelRehim:2005gz}.) Again, GMRES(20) fails to converge. The sharp knee in Fig.~1 signaling deflation is replaced by a more gradual slope change. In this case there is only a small change in the convergence properties in going from 30 to 40 deflated eigenvectors, after which there is saturation. This problem takes more iterations to converge than the Wilson one, and the gain from deflation (GMRES-Proj relative to GMRES-DR) is not as dramatic, but it still helps considerably. We will see later that the dynamical twisted mass case is also a tougher problem than the Wilson/clover one in terms of the number of iterations rquired.

We also consider matrix shifting for the GMRES-DR, GMRES-Proj algorithms, also called \lq\lq multi-mass" because the shifts represents different quark masses in the Wilson formulation. Krylov subspaces are shift invariant in that $A-\sigma_i I$, where $I$ is the identity matrix and $\sigma_i$ are numbers, generates the same Krylov subspace as $A$ no matter what the shift. Given this, the goal is to solve all shifted systems with this single Krylov subspace. For non-restarted methods this has been done, for example, for QMR by Freund~\cite{Freund} and BiCGStab by Frommer~\cite{Frommer}.

Restarting makes the situation more difficult because the shifted residuals are not parallel to one another, generating different Krylov subspaces. Frommer and Glassner, in the context of restarted GMRES~\cite{FandG} have provided a solution: force all residuals to all be parallel after a restart. One can then continue using a single Krylov subspace for all shifted systems. However, the minimal residual property is maintained only for the base shift system. In spite of this, we have not found this to be a problem. The residuals for the shifted systems for QCD problems are often indistinguishable from a non-shifted run at the same mass.

The subspaces generated by GMRES-DR are a combination of approximate eigenvectors portion and Krylov subspace portion (see (\ref{ss}) above), but as pointed out above, they are Krylov themselves. Therefore, GMRES-DR can be restarted like GMRES for multiple shifts.

Deflating eigenvalues is difficult for multiple shifts because one can not keep the residual vectors parallel unless one has exact eigenvectors. Lack of exact eigenvectors will cause the additional right-hand sides to \lq\lq stall out". The solution is to force the error to be in the direction of one vector, namely $v_{k+1}$ from Eq.(\ref{AV}). One can then can correct the error at the end. One needs the solution of one extra right-hand side to accomplish this, but this cost will be low for many right-hand sides. The shifted version of GMRES-DR($m,k$) is called GMRES-DRS($m,k$) and the shifted version of GMRES($m$)-Proj($k$) is called GMRES($m$)-Proj($k$)-Sh.

To show details on the method, consider Fig.~4. This uses a non-hermitain, bidiagonal matrix of order 2000 for illustration purposes, and is solved in MATLAB. The eigenvalue spectrum is $0.1,1,2,3,\dots, 1999$. The left part of the figure shows the solution of the first right-hand side, using GMRES-DRS(25,10). The solid blue line is the base system and the red dotted line is the shifted one. The effect of deflation is seen again with the sharp knee in the residual norm. The next pair of falling lines represent the solution of the extra right-hand side, which is solved to lower accuracy. Next the second right-hand side is solved with GMRES(15)-Proj(10)-Sh. The green dotted line shows the uncorrected residual norm for the shifted system, which stalls out at $\approx 4.0 \times 10^{-3}$, while the green dash-dot line shows the second shift residuals as if they were corrected. Actually, the correction needs to be done only once at the end of each right-hand side! The new right-hand sides were only converged to $\approx 10^{-5}$, but that can easily be changed by running the base/shifted system longer. Notice the convergence is faster for the additional right-hand sides than for GMRES-DRS, because the eigenvectors are used from the beginning to speed up the convergence. Also the cost of GMRES(15)-Proj(10)-Sh is less than for GMRES-DRS(25,10), because it is fairly inexpensive to project over the approximate eigenvectors compared to keeping the eigenvectors in the GMRES-DR subspace.

\begin{figure}
\begin{center}
\leavevmode
\includegraphics*[viewport=0 0 800 520, scale=0.46]{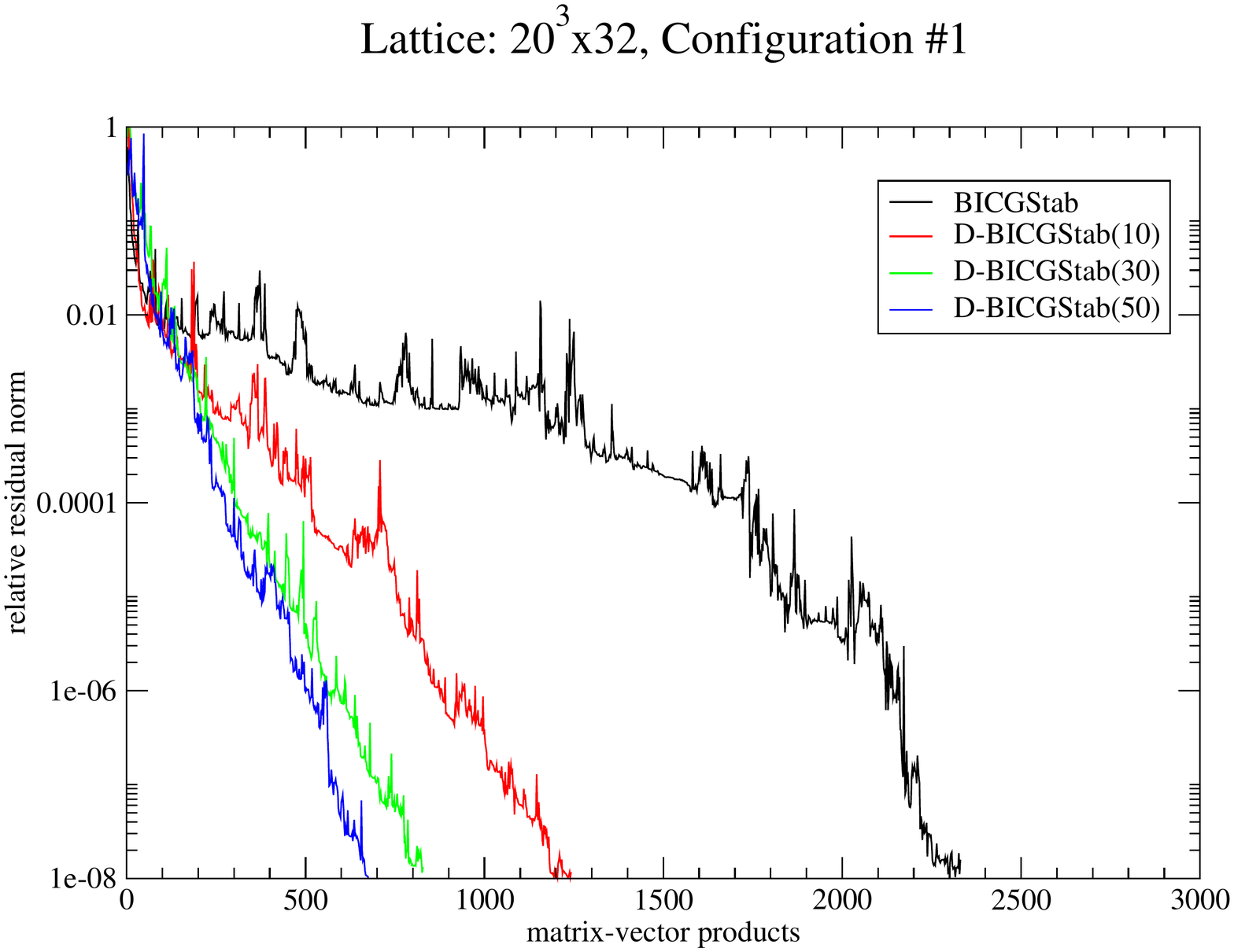}
\caption{The residual norm as a function of matrix vector products on a $20^3\times 32$ Wilson lattice. Regular BiCGStab is compared to D-BiCGStab($k$) for $k=$10, 20 or 30 low eigenvectors on the same lattice and right-hand side.}
\end{center}
\end{figure}

\begin{figure}
\begin{center}
\leavevmode
\includegraphics*[viewport=0 0 800 520, scale=0.46]{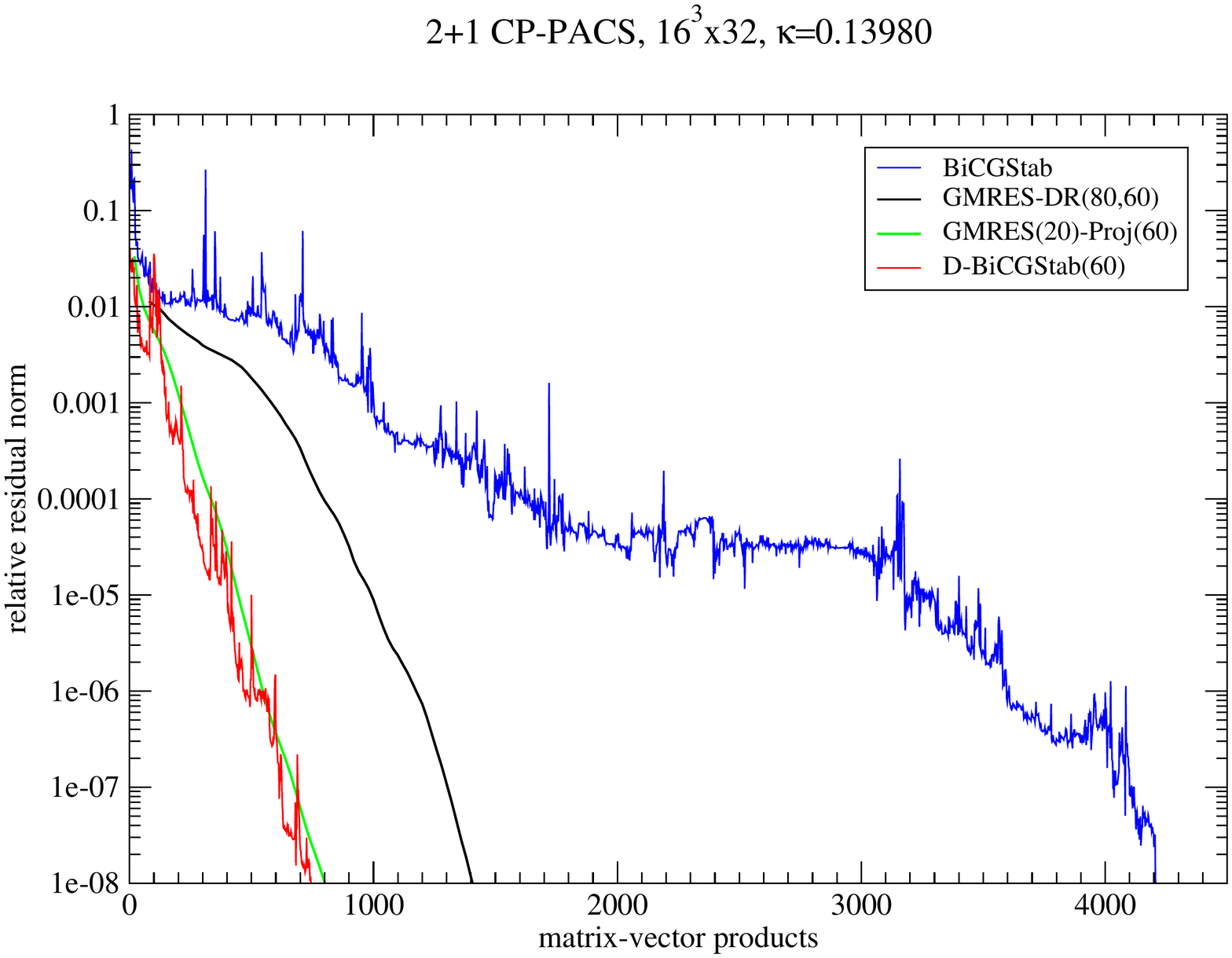}
\caption{The residual norm as a function of matrix vector products on a $16^3\times 32$ dynamical 2+1 CP-PACS~\cite{CP-PACS} configuration, evaluated at $\kappa=0.13980$.}
\vspace{1cm}
\includegraphics*[viewport=0 0 800 510, scale=0.46]{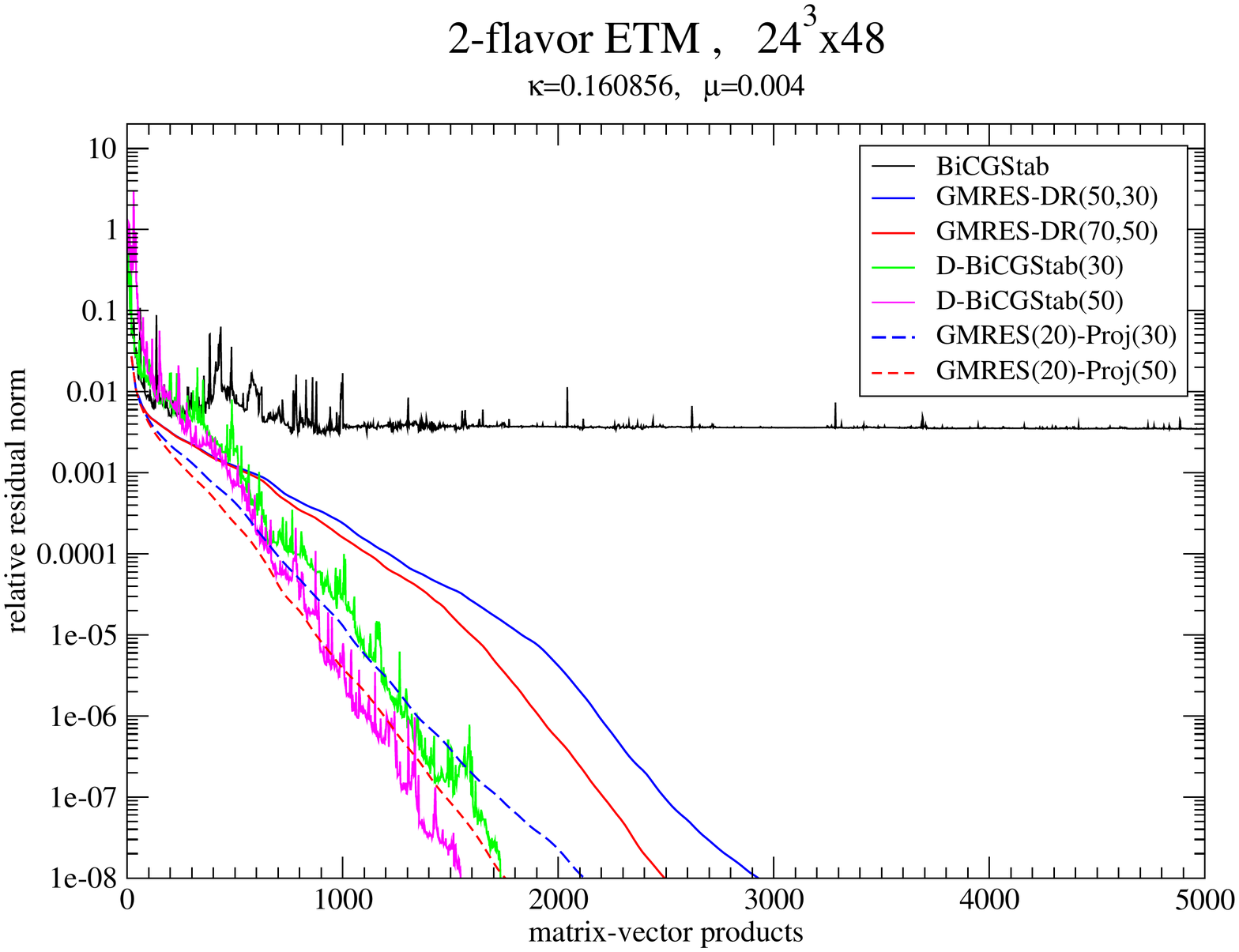}
\caption{The residual norm as a function of matrix vector products on a $24^3\times 48$ dynamical twisted mass configuration from the ETMC~\cite{ETMC}, evaluated at $\kappa= 0.160859$ and $\mu=0.004$, the smallest quark mass used in this work.}
\end{center}
\end{figure}

Doing multiple shifts while deflating using GMRES-DRS and GMRES-Proj-Sh is illustrated in Fig.~5 for a $16^4$ Wilson matrix at $\kappa=0.158$. This is approximately $\kappa_{cr}$. Here we just show the solution of the second right-hand side for the base system and two shifts, leaving out the solution of the extra right-hand side. The first right-hand side is solved with GMRES-DR(50,30) to residual tolerance of $10^{-8}$ and the extra right-hand side to $10^{-7}$. For the second right-hand side, GMRES-Proj uses 30 approximate eigenvectors for the projection between cycles of GMRES(20). GMRES(20)-Proj(30)-Sh converges in about one-tenth of the iterations needed for GMRES(20). To reach a residual norm of less than $10^{-7}$ for the base (unshifted) system takes 2680 matrix-vector products for GMRES(20)-Sh and 280 for GMRES(20)-Proj(30)-Sh.

The GMRES methods are all restarted, and it is necessary to keep an associated Krylov set of vectors in memory. Non-restarted methods for large matrices often converge faster than restarted GMRES. For those who prefer not to restart but still want to solve multiple right-hand sides we have developed a deflated BiCGStab (\lq\lq D-BiCGStab($k$)") which deflates $k$ eigenvectors. This has been described previously in Refs.~\cite{NP2} and \cite{realfirst}. The problem is that a simple projection over eigenvectors is not good enough to last for the entire run of BiCGStab. Our solution is to use a projection over both right and left approximate eigenvectors. We will see that this is very effective, and in addition, the left eigenvectors may be obtained from the right eigenvectors for Wilson fermions using the $(M^{-1})^{\dagger}=\gamma_5 M^{-1}\gamma_5$ identity.

Fig.~6 shows the effects of keeping different numbers of deflated eigenvectors on the residual norms as a function of MVPs during runs of D-BiCGStab($k$) on a $20^3\times 32$ Wilson lattice. Deflating about 50 eigenvectors on the $20^3\times 32$ seems to be optimal, after which residual norm saturation sets in. The typical speedup on a lattice of size $16^4$ is about 2.7 compared to regular BiCGStab, while the $20^3\times 32$ lattice speedup increases to about 5. Occasionally, an additional \lq\lq breakthrough" speedup occurs; see examples in our PoS proceedings paper~\cite{PoS1}. Again, these runs are done very close to $\kappa_{cr}$.

Figs.~7 and 8 compare the effect of GMRES-Proj and D-BiCGStab on dynamical configurations obtained from CP-PACS~\cite{CP-PACS} and the ETMC~\cite{ETMC}, respectively. Notice the slow curvature of the residual curve for GMRES-DR(80,60) in Fig.~7, as opposed to the sharp knee of deflation in Fig.~1 for quenched Wilson fermions. This implies an almost continuous deflation of eigenvalues during the run. The final slope of the residual norm is significantly larger than the initial and this slope is applied essentially from the beginning of the GMRES(20)-Proj(60) or the D-BiCGStab(60) runs leading to a significant speedup. As we usually find, the non-restarted D-BiCGStab beats the GMRES-Proj at the end. Fig.~8 shows an even tougher problem, the lowest twisted mass from the ETM Collaboration. Regular BiCGStab does not converge. Two GMRES-DR curves are shown, for 30 and 50 eigenvectors, as well as corresponding D-BiCGStab runs. The superlinear effect of deflation, compared to dynamical Wilson/clover, is not as noticeable, just as for the quenched case. Again, D-BiCGStab slightly beats GMRES-Proj to full convergence. Note that for the D-BiCGStab run we had to prepare both left and right eigenvectors, which are separate calculations for the twisted case, as opposed to Wilson. However, the \lq\lq down" eigenvectors are the \lq\lq up" left eigenvectors, aside of a $\gamma_5$ factor, and vice versa. This means there is no extra work involved for left eigenmode preparation if both up and down quark propagators are always calculated.

The optimal number of eigenvalues grows slowly with the size of the problem. On quenched lattice problems we have found that it increases from about 10 on small lattices ($8^4$) to about 40 to 50 on the largest lattices we have examined ($20^3\times 32$), a scale change of over 60. Another very important question concerns the iteration count of the algorithm. If the number of iterations grows as the space-time volume of the lattice, $V$, then the number of computations is proportional to $V^2$, given that the length of MVPs, which dominate the expense, is also proportional to $V$. We will call this the algorithm acceleration problem~\footnote{L\"uscher states a very specific way algorithm growth occurs during a QCD deflation process and calls it the \lq\lq $V^2$ problem". In contrast, I do not take a stand on how the algorithm growth comes about, but simply examine the growth in the number of the algorithm's algebraic steps during \lq\lq peak" speedup.}. Accordingly, after eigenvalues are deflated, how does the number of iterations vary with the size of the problem? We find that the average number of iterations in going from the $16^4$ to the $20^3\times 32$ lattice (a volume change of 3.9) increased from about 300 to about 600 for GMRES-Proj and from about 300 to about 500 iterations for D-BiCGStab. Using the better results from D-BiCGStab and scaling the change in volume to 3 in order to compare with the other two algorithms, this represents an approximate 45\% increase in the number of iterations. For comparison, the corresponding BiCGStab increase for this change is about $100\%$.

\section{Stathopoulos' and Orginos' work on CG deflation}

\begin{figure}
\begin{center}
\leavevmode
\includegraphics*[viewport=0 0 600 420, scale=0.60]{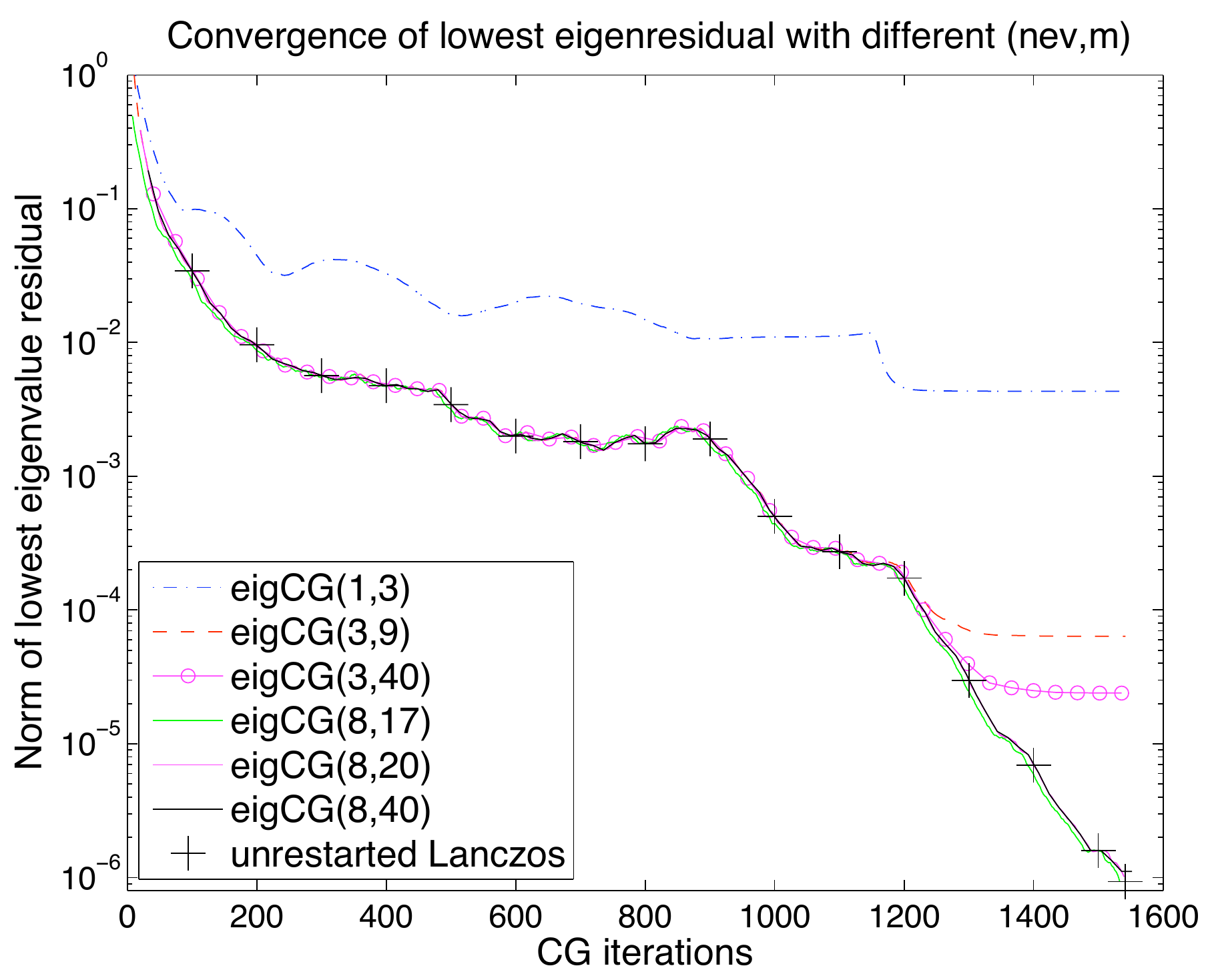}
\caption{The residual norm of the lowest extracted eigenvector of the Stathopoulos/Orginos algorithm  compared to unrestarted Lanczos for a small Wilson lattice.}
\end{center}
\end{figure}

There are a number of problems in lattice QCD for which a non-hermitian approach is not necessary and possibly inefficient. The shifted inner loop of overlap fermions is one, the $\gamma_5$-hermitian Wilson-Dirac operator is another, and, if one is willing to use the normal equations, the Wilson/clover matrix is another. The optimal method is Conjugate Gradient (CG) for hermitian systems. Stathopoulos  and Orginos have recently published a deflation algorithm using CG~\cite{lbl12}.

Their algorithm comes in three parts. The first part, eig-CG($nev,m$), like GMRES-DR, solves linear equations and does simultaneous improvements of the deflated eigenvectors. Although the eigenvector calculation part of their algorithm is restarted, it does not affect the solution of the CG linear equations, which are not restarted, only interrupted. The second part, incremental eig-CG($s$), where $s$ labels the right-hand sides, can be thought of as a controller for eig-CG. It calls eig-CG, adds $nev$ new eigenvectors to a separate, cumulative eigenvector subspace after each right-hand side, and does a Rayleigh-Ritz orthogonalization of the full set. It is used for the first $s\le s_1$ right-hand sides. The third part, init-CG, is simply standard CG with an initial (and one additional) Galerkin projection using the completed set of eigenvectors for the final $s>s_1$ right-hand sides.

\begin{figure}
\begin{center}
\leavevmode
\includegraphics*[viewport=0 0 600 385, scale=0.50]{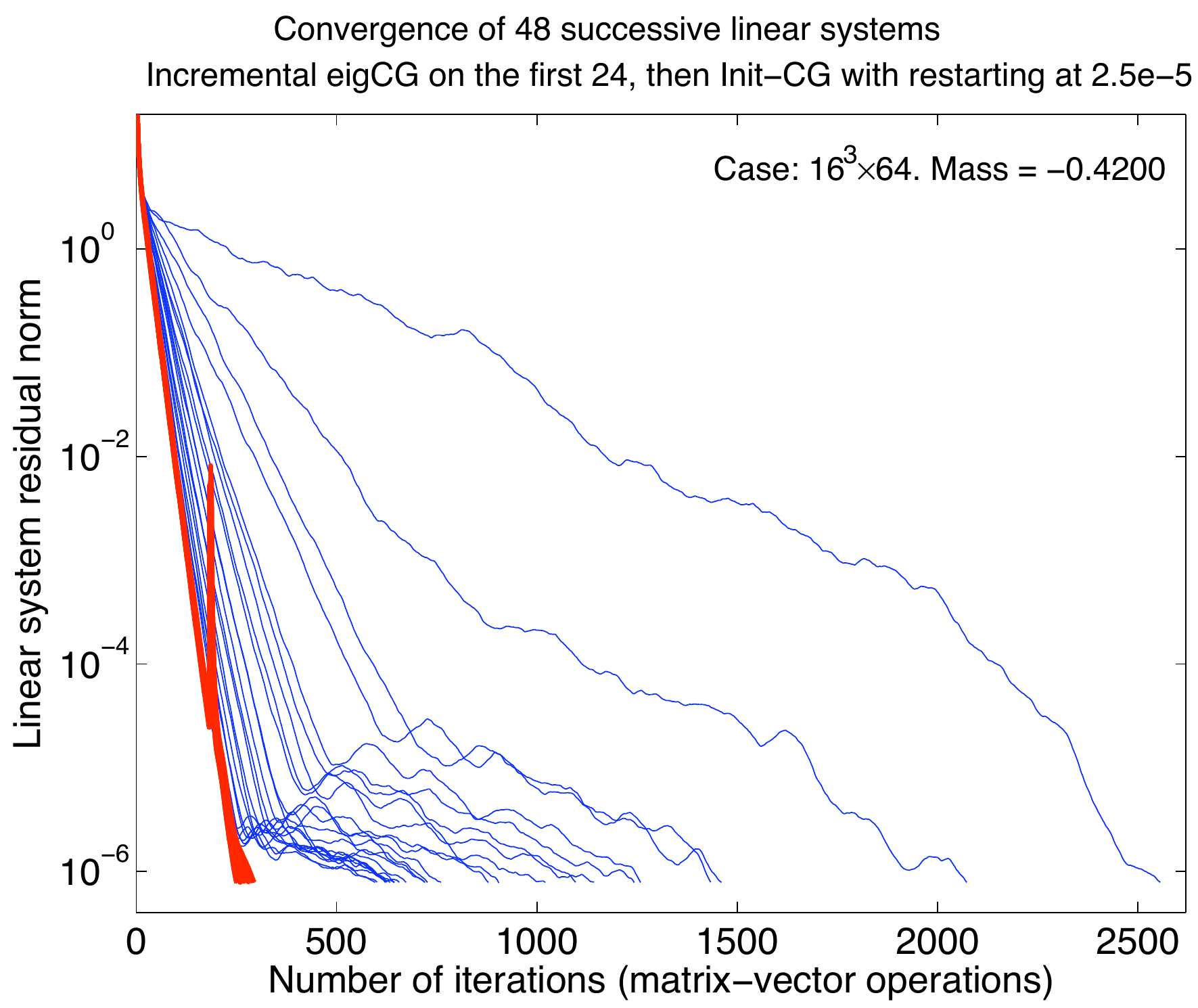}
\vspace{1cm}

\includegraphics*[viewport=0 0 600 385, scale=0.50]{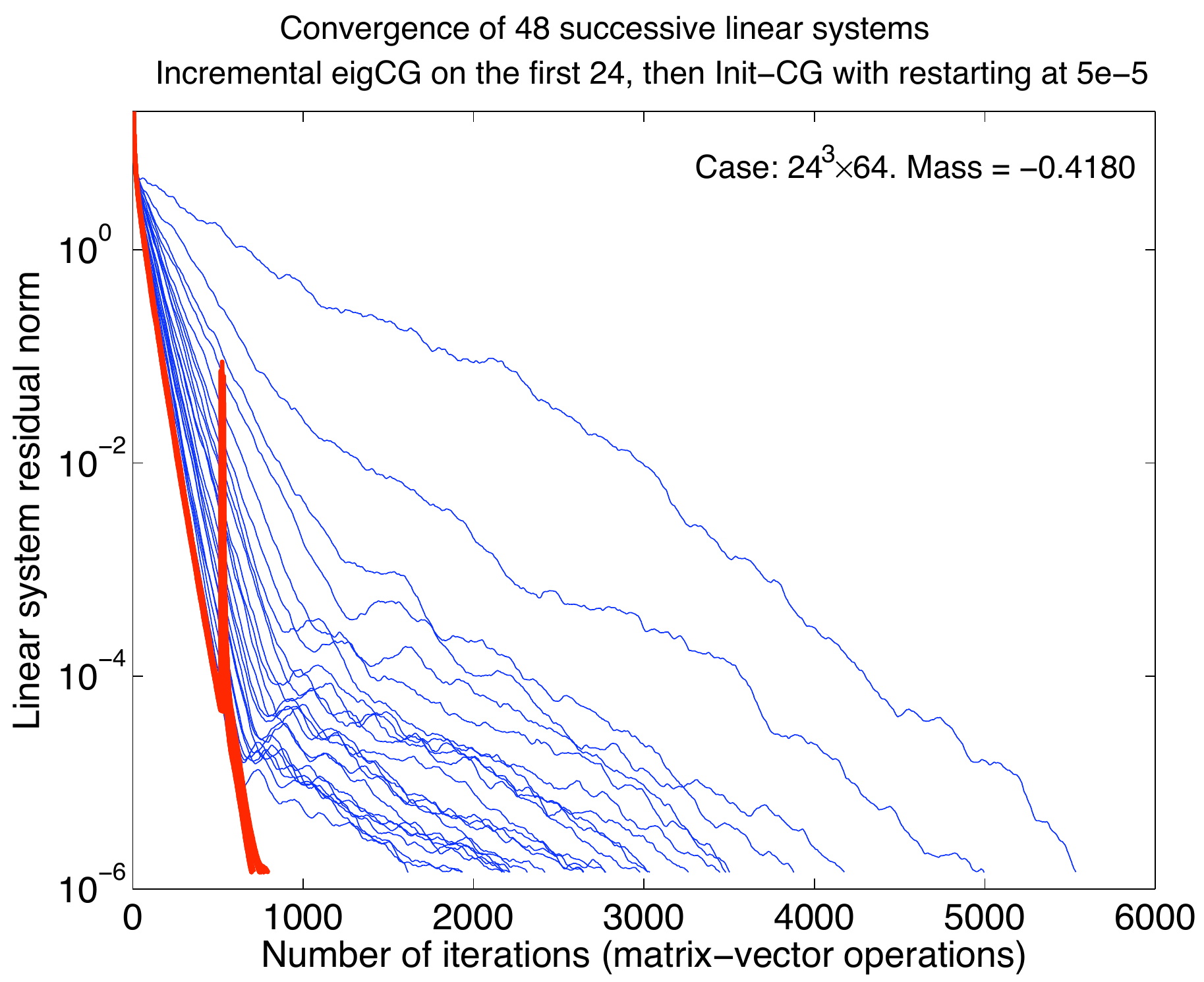}
\caption{The convergence history of the Statopoulos and Orginos lattice fermion solver for the $16^3\times 64$ lattices (upper figure) and $24^3\times 64$ lattices (lower figure) near $m_{cr}$.}
\end{center}
\end{figure}

\begin{figure}
\begin{center}
\leavevmode
\includegraphics*[viewport=0 0 600 400, scale=0.50]{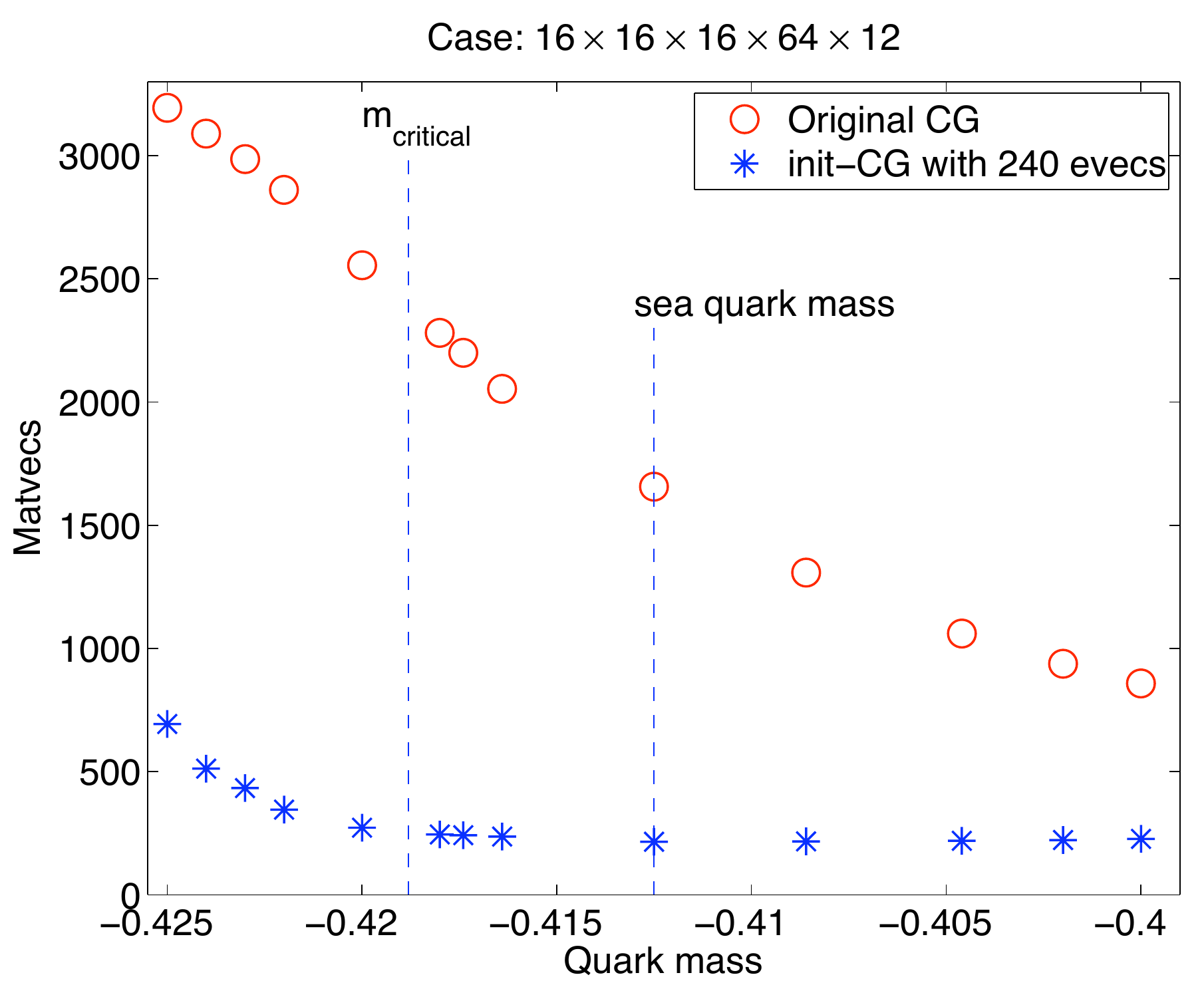}

\vspace{1cm}

\includegraphics*[viewport=0 0 600 400, scale=0.50]{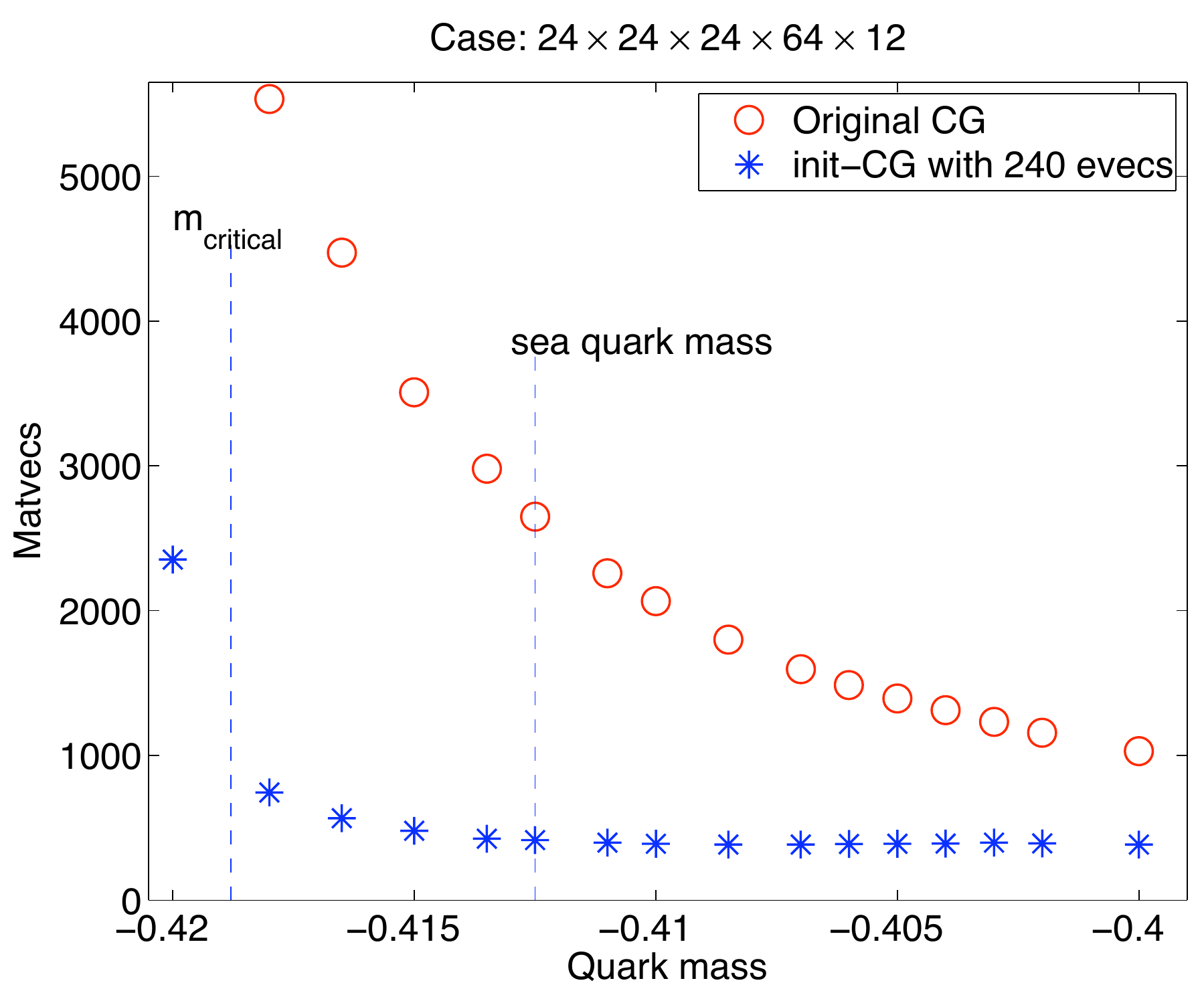}
\caption{The quark mass dependence of the matrix vector operations of the Statopoulos and Orginos lattice fermion solver for the last 24 right-hand sides compared with the undeflated CG solve for the $16^3\times 64$ lattices (upper figure) and $24^3\times 64$ lattices (lower figure).}
\end{center}
\end{figure}

In some more detail, eig-CG($nev,m$) has a restarted subspace of maximum dimension $m$, which is made up of $nev$ previous eigenvectors, $nev$ current eigenvectors and $(m-2)*nev$ Krylov vectors. It uses Rayleigh-Ritz to compute the low eigenvectors and appends a portion of the CG search space to the eigenvectors. Typically, however, the linear equations converge faster than the eigenvalue part. Thus, something additional needs to be done to assure convergence, since eigenvalue accuracy is crucial to deflation. Incremental eig-CG($s$) ($s = 2,\dots$) keeps track of the $(s-1)*nev$ previous eigenvectors, calls eig-CG, and accumulates another $nev$ approximate Ritz vectors from each new right hand side, $s$, for $s<s_1$. In the configuration run by Stathopoulos and Orginos, it needs significant storage.

Stathopoulos and Orginos tested this algorithm on several anisotropic, 2 flavor Wilson fermion gauge fields, using the normal equations to solve for the hermitian combination $M^{\dagger}M$, where $M$ is the Wilson-Dirac operator. The number of Wilson/Dirac MVPs (which are even/odd preconditioned) is therefore twice the number of the algorithm's iterations. There is an anisotropy factor in the time direction of 3 for their lattices. As tested, their algorithm uses single precision, except on dot products where double precision is used. They chose the maximum subspace size to be $m=100$. In addition, $nev=10$, which means that a restart cycle (after the first) contains $100-2*10=80$ iterations. They also tested for 48 total right-hand sides with $s_1=24$. This was imlemented within Chroma, the lattice QCD C++ based software developed at Jefferson National Lab~\cite{chroma}.

An interesting point that emerges in their numerical solves is that in spite of the fact that they restart the eigenvector part of their algorithm, it seems to converge as fast as unrestarted Lanczos. This is illustrated in Fig.~9. They ran various configurations of eig-CG($nev,m$) on a smaller Wilson matrix. When run with a small number of eigenvectors, the accuracy stagnates. However, keeping only 8 eigenvectors is sufficient to track unrestarted Lanczos.

Fig.~10 gives the convergence history of the linear equations (not the eigenvectors) for the $16^3\times 64$ and $24^3\times 64$ lattices, both near $m_{cr}$, as a function of iteration count. Blue lines represent the first 24 right-hand sides using incremental eig-CG (calling eig-CG), the red lines the last 24 using init-CG. Note the sub-linear convergence which occurs near single precision machine accuracy, $10^{-5}$, which develops as more and more right hand sides are introduced. However, also notice the concomitant incremental decrease in the number of iterations necessary to solve each system. For the red lines, representing the last 24 init-CG runs, also note the quickly disappearing spike at the single restart, which is done as the systems again approach machine accuracy.

The blue stars in Fig.~11 give the quark mass dependence of the iteration count needed for convergence on the last 24 right-hand sides. This figure also shows the same runs done with non-deflated CG with red circles. By comparing the approximate values given in this graph, one can deduce the approximate speedup compared to the original CG. The peak speedup is $\approx 10$ on smaller lattice near $m_{cr}$, whereas the approximate integrated speedup, using Fig.~10, is $\approx 4$ (all right-hand sides). For the larger lattices, the peak speedup is $\approx 7$ on larger lattice near $m_{cr}$, and the integrated value is near $3$.

Another important aspect concerns the increase in the number of iterations of the algorithm for a given increase in the lattice volume. The two lattices examined, $16^3\times 64$ and $24^3\times 64$, have an approximate volume change of about $3.4$. Fig.~11 shows that the iteration number on the last 24 right-hand sides changes from about 250 on the smaller lattices to 800 on the larger near $m_{cr}$. Scaling the iteration number increase to a volume increase factor of $3$ instead of $3.4$, this amounts to an approximate $190\%$ increase in iterations, which is substantial. (The corresponding increase in the number of iterations of the original CG is only $140\%$.) The algorithm would probably do better on the larger lattice if a larger number of eigenvectors were deflated, but this would increase the orthogonalization cost and possibly make the problem so large that it would no longer fit in computer memory. Also, double precision would help defeat the sub-linear convergence. Nevertheless, the flat behavior in Fig.~11 across quark masses and the peak speedup on these lattice is impressive.

\section{L\"uscher's work on domain decomposition deflation using GCR}

L\"uscher's domain decomposition algorithm introduces an interesting new type of deflation for lattice type systems. This algorithm breaks the lattice problem up into an \lq\lq inner" deflation subspace, $S$ (defined on $4^4$ blocks, of dimension $N$, defined later) and the \lq\lq outer" orthogonal complement, $S^{\perp}$. The basic deflated system is described as
\begin{equation}
\psi(x)=\chi(x)+\sum_{k.l}^{N}{\phi_k (A^{-1})_{kl}}(\phi_l,\eta),
\end{equation}
where
\begin{equation}
P_LD\chi(x)=P_L\eta(x), (1-P_R)\chi(x)=0.
\end{equation}
\lq\lq $D$" is the Wilson-Dirac/clover operator, the $\phi_l$ are basis fields on the blocks, and $A_{kl}$ is the matrix representing the Wilson-Dirac/clover operator on those blocks:
\begin{equation}
A_{kl}=(\phi_k,D\phi_l).
\end{equation}
The definition of the projection operators $P_L$ and $P_R$ is
\begin{eqnarray}
P_L\psi(x)=\psi(x)-\sum_{k.l}^{N}{D\phi_k (A^{-1})_{kl}}(\phi_l,\psi),\\
P_R\psi(x)=\psi(x)-\sum_{k.l}^{N}{\phi_k (A^{-1})_{kl}}(\phi_l,D\psi).
\end{eqnarray}
The outer part of L\"uscher's algorithm uses the Schwartz Alternating Procedure (SAP) matrix~\cite{SAP,Lusch3} as a right preconditioner on $4^3\times 8$ blocks and the Generalized Conjugate Residual (GCR)~\cite{Saad} algorithm for the Krylov inverter on the $S^{\perp}$ space. After SAP preconditioning, the basic solve involves
\begin{eqnarray}
P_LDM_{SAP}\phi(x)=P_L\eta(x),\\
\chi(x)=P_RM_{SAP}\phi(x).
\end{eqnarray}
Please see Refs.~\cite{Vuik1,Vuik2} for more mathematical background on preconditioning and deflation on domains. 

\begin{figure}
\begin{center}
\leavevmode
\includegraphics[scale=0.72]{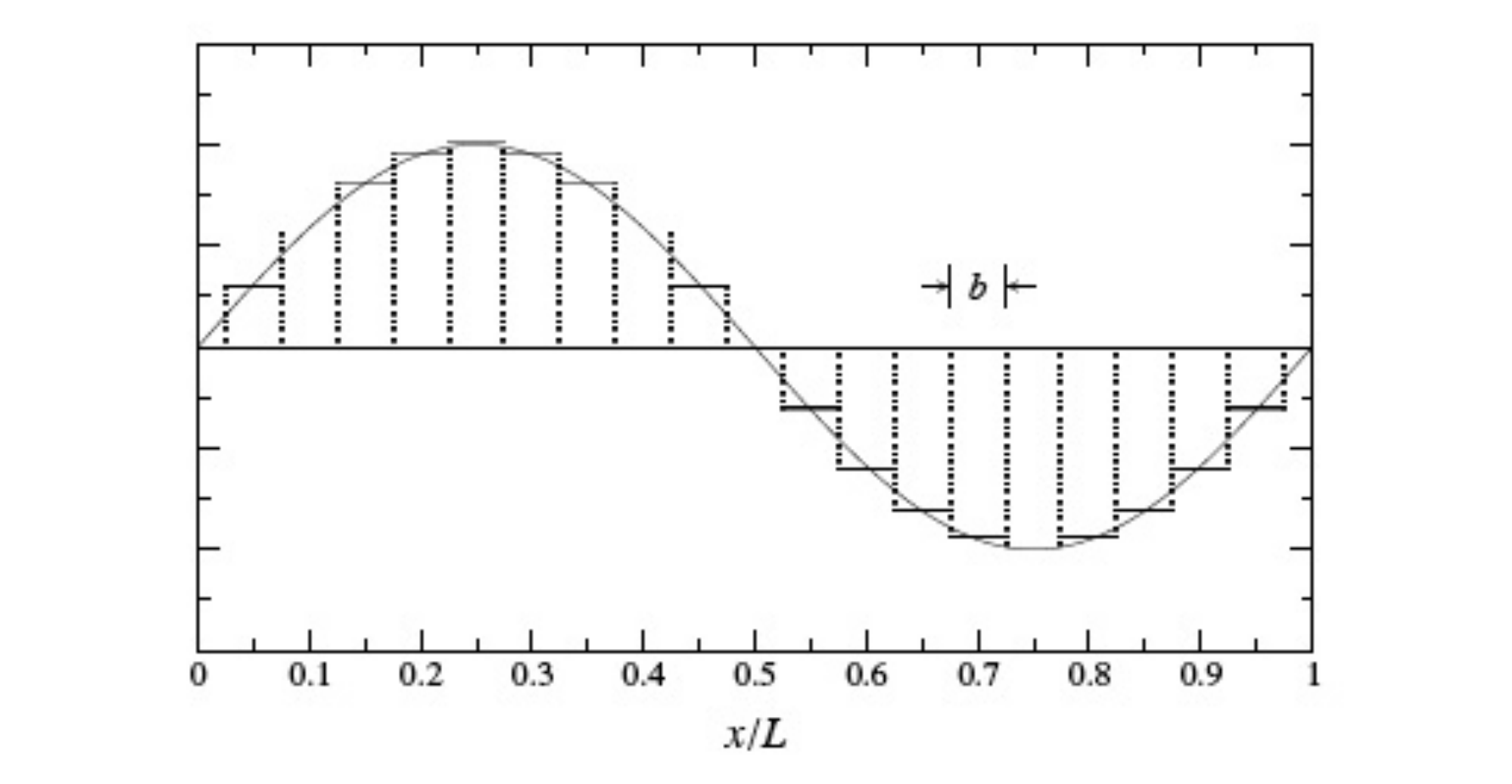}
\caption{Figure pertaining to the idea of \lq\lq local coherence" within the context of domain decomposition. Low eigenmodes for free quarks on a periodic lattice are projected out using constant fields on domains.}
\vspace{2cm}
\includegraphics[scale=0.70]{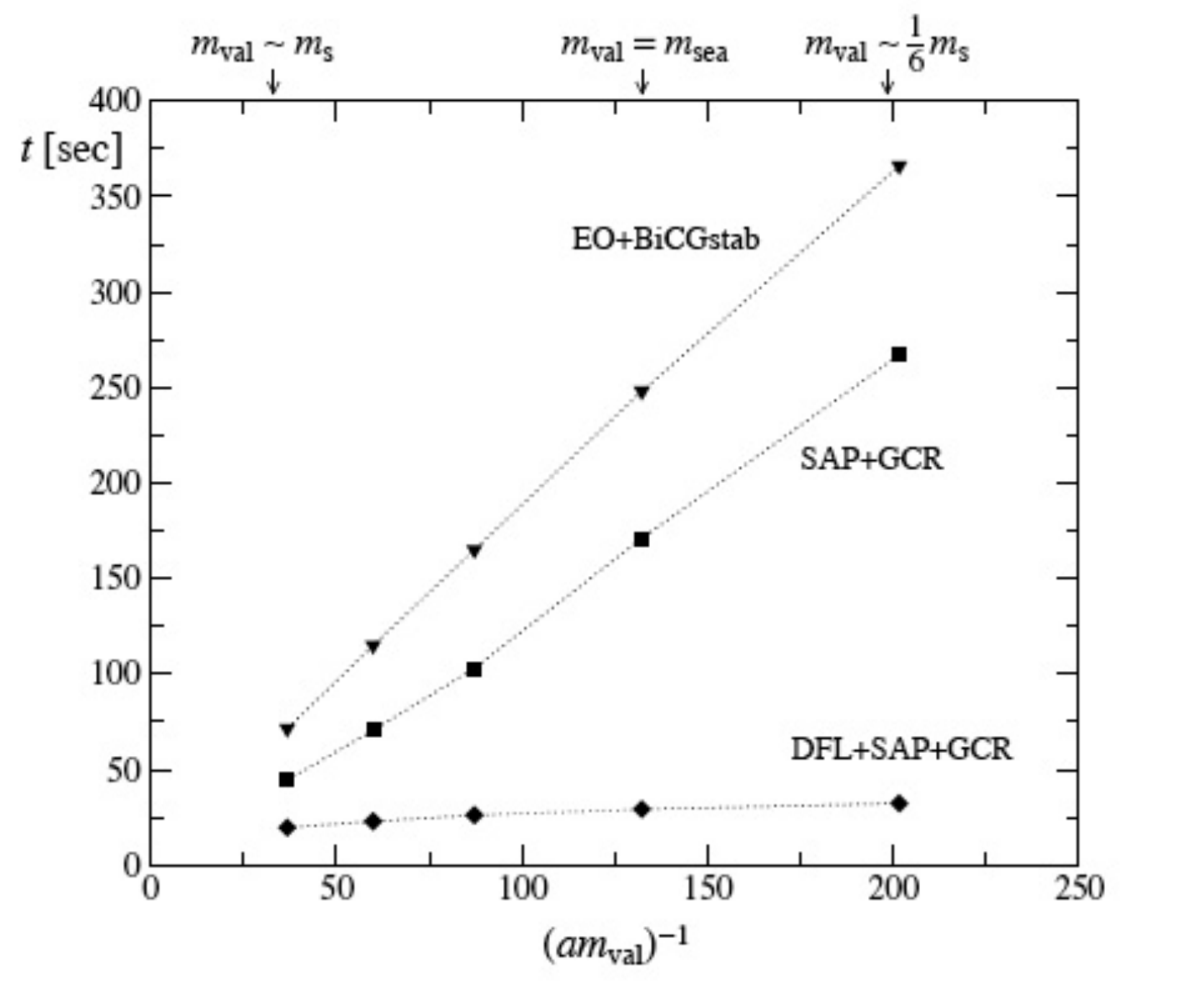}
\caption{Comparison of the solver times for even/odd preditioned BiCGStab (\lq\lq EO+BiCGStab"), domain decomposition (\lq\lq SAP+GCR"), and deflated domain decomposition (\lq\lq DFL+SAP+GCR") on a  $32^3\times 64$ lattice.}
\end{center}
\end{figure}

The algorithm actually starts with subspace generation for later use in deflation. It applies inverse iteration (in the form of SAP) to $N_s$ (chosen to be 20) random global vectors to get linear combinations of low eigenmodes, $\psi_l$, which are then projected onto the domains and orthonormalized. These are the block fields, $\phi_l$. The subspace is prepared a single time near $m_{cr}$ and does not need to be repeated.

The algorithm solves the Wilson-Dirac/clover matrix on the full system knocking out the deflated eigenvectors with the $P_L$ projector while using the GCR algorithm. It considers one quark mass at a time. GCR allows for inexact preconditioning and works well with SAP. Preconditioning is important because it reduces the iteration count of GCR and therefore the deflation overhead. The dimensionality, $N=N_sN_b$, where $N_b$ is the number of blocks, of the inner part of the algorithm is fairly large. It is of size $20\times 2592=51,840$ for the smaller lattice ($24^3\times 48$) and $20\times 8192=163,840$ for the larger ($32^3\times 48$). It can not be solved exactly for a given source vector, as one does on a typical Rayleigh-Ritz deflation subspace, but must be solved iteratively. It is even-odd preconditioned and solved with GCR. This inner space also is deflated, using the global vectors, $\psi_l$, used to prepare the deflation subspace.

Deflation on domains works because of \lq\lq local coherence". Only a small number of global vectors projected onto the blocks are needed to project out low modes. This is analogous to filtering out low modes for free quarks on a periodic lattice. L\"uscher illustrates this in Fig.~12, where constant fields on the domains are used to approximately project the low, smooth fields. It illustrates that high deflation efficiencies can be achieved by subspaces of fields that are only piecewise smooth and are not approximate eigenmodes of the Dirac operator themselves.

 L\"uscher has done his development work within the context of dynamical 2 flavor Wilson/clover fermions (50 configurations). Fig.~13 gives the final results of the tests, outside of the overhead for subspace generation. Comparison is made between even/odd preconditioned BiCGStab, domain decomposition, and domain decomposition deflation on the larger of the two lattices. This figure gives the average execution time for the solution of the system as a function of inverse quark mass in lattice units. The dependences are approximately linear for all three algorithms. The total overhead for subspace generation was 150s for the $24^3\times 48$ lattice and 184s for $32^3\times 64$ lattice (tuned to $m_{cr}$). These overheads can not be compared to each other directly because they are done on different computer arrays. Using the Table 1 numbers from Ref.~\cite{Lusch}, the peak speedup of the algorithm at the lightest quark mass, relative to BiCGStab on the same system, is $366/32=11.4$. The integrated speedup relative to BiCGStab may be computed in various ways depending on usage. For just the 5 mass solves used in Table 1 on the large lattice, one obtains $966/314=3.1$ when the overhead of 184s is included. A more typical usage would be to compute at least 1 quark propagator (= 12 solves) at each mass. Considering all preparatory work, including the initial inner Dirac solve, the integrated speedup factor is then (12*966)/(12*130+184+4*3.9)=6.6.

The volume change in going from the small ($24^3\times 48$) to large lattices ($32^3\times 48$) is about $3.2$. Scaling this down to a volume change of 3 again, L\"uscher's Table 1 shows there is an approximate $13\%$ increase in the number of outer GCR iterations for this change. L\"uscher's Table 2 also shows there is an accompanying increase of approximately $45\%$ in the inner GCR iterations for the same volume change. Although this constitutes only a fraction of the time spent for the solution of the full system ($20-30\%$), this fraction will increase for larger lattices. For comparison to the deflated algorithm's $13\%$ increase in the number of outer iterations, note that the increase in the number of BiCGStab iterations for the same scale change is about $30\%$.

The iterative growth of the Morgan/Wilcox and L\"uscher algorithms, relative to BiCGStab, seem to be very close. For the scale change of three, the iteration change ratio for Morgan/Wilcox is $45\%/100\%\approx 0.45$, and for L\"uscher it is $13\%/30\%\approx 0.43$. Of course these come from very different sized lattices and the Morgan/Wilcox lattices are quenched.

\vspace{.5cm}
\section{Comparison table}

\begin{table}[htbp]
\begin{center}
\begin{tabular}{|c|c|c|c|}
\hline
&{\bf Morgan/Wilcox} & {\bf Stath./Orginos} & {\bf L\"uscher}\\
\hline\hline
{\bf Basic solvers} & GMRES; BiCGStab & CG & GCR\\\hline
{\bf Matrix type} & non-hermitian & hermitian & non-hermitian \\
&(algebraic)&(algebraic)&(lattice)\\\hline
{\bf Preconditioning?} & no & no & yes (SAP)\\\hline
{\bf Simultaneous solve?} &yes&yes&no\\\hline
{\bf Eigenvalue use} & Every cycle (GMRES-Proj)& Every cycle ($s\le s_1$); & Every outer\\
{\bf for multiple rhs's} &At beginning (BiCGStab)&beginning &iteration \\
&&and restart ($s>s_1$)& \\\hline
{\bf Matrix shifting?} &yes (GMRES) & no & no \\
&no (BiCGStab)&&\\\hline
{\bf Algorithm } &mild & large & small\\
{\bf acceleration}&&&\\
\hline
\end{tabular}
\caption{Some points of comparison for the three algorithms considered.}
\end{center}
\end{table}

For purposes of quick comparison of the three algorithms outlined above, Table~1 may be useful. The columns list the three algorithms and the rows give some of their major aspects. Some explanation is necessary. The \lq\lq Matrix type" entry is pointing out that the Morgan/Wilcox and Stathopoulos/Orginos algorithms are general algebraic solvers, whereas L\"uscher's domain decomposition method works on a lattice-defined system. Although for the preconditioning question I answer in the negative for the Morgan/Wilcox and Stathopoulos/Orginos cases, both use an even/odd solve for the problem addressed. (L\"uscher also uses such a preconditioning for his \lq\lq inner" Dirac solve.) The \lq\lq Simultaneous solve?" row concerns the question of whether the deflation steps are taking place simultaneously with the solution of the linear equations. For the domain decomposition algorithm this is a separate step, which, however, does not need to be repeated for the different quark masses. The \lq\lq Eigenvalue use for multiple rhs's" entry is giving the frequency with which eigenvectors or linear combinations of eigenvectors are projected within the algorithm. For the Stathopoulos/Orginos algorithm this is done cyclically (cycle chosen is 80 iterations) for the first $s_1$ right-hand sides, but are only used at the beginning (and one restart) for the remainder. For L\"uscher's domain decomposition actually both the inner and outer GCR loops are deflated. The \lq\lq Algorithm acceleration" row concerns how the iteration count increases for an increase in the size of the problem. We have chosen to list the Morgan/Wilcox increase as \lq\lq mild", but a true comparison with the matrices used by L\"uscher should be done for the same size matrices, where we would expect a continued increase in the effectiveness of our algorithms while maintaining slow eigenvalue number growth.

\section{Acknowledgments}

The author thanks the Sabbatical Committee of the College of Arts and Sciences of Baylor University for support. The author's work was done on the Baylor University High Performance Cluster as well as the Mercury Cluster at the National Center for Supercomputing Applications (NCSA). The author would especially like to thank Dr.~Ron Morgan of the Baylor Mathematics Department with whom this joint work on lattice QCD applications was carried out, as well as Dr.~Abdou Abdel-Rehim, a Baylor University Postdoctoral Fellow. In addition, we thank the CP-PACS Collaboration as well as the ETM Collaboration for the use of their lattices.

\end{document}